\documentstyle[epsfig,12pt,a4p]{article}
\newcommand{\be}{\begin{equation}} \newcommand{\ee}{\end {equation}}

\newcommand{\pom}{ $I\hspace{-1.6mm}P$}
\begin{document}
\bibliographystyle{unsrt}
\def\question#1{{{\marginpar{\small \sc #1}}}}

\rightline{29 June 2000}
\baselineskip=18pt
\vskip 0.7in
\begin{center}
{{\bf \LARGE Dynamics of Glueball and $q\bar{q}$ production in the
central region of pp collisions}}\\
\vspace*{0.9in}
{\large Frank E. Close}\footnote{\tt{e-mail: F.E.Close@rl.ac.uk}} \\
\vspace{.1in}
{\it CERN, Geneva, Switzerland}\\
{\it and}\\
{\it Rutherford Appleton Laboratory}\\
{\it Chilton, Didcot, OX11 0QX, England}\\
\vspace{0.1in}
{\large Andrew Kirk}\footnote{\tt{e-mail: ak@hep.ph.bham.ac.uk}} \\
{\it School of Physics and Space Research}\\
{\it Birmingham University}\\
\vspace{0.1in}
{\large Gerhard Schuler}\footnote{\tt{e-mail: Gerhard.Schuler@cern.ch}} \\
{\it CERN, Geneva, Switzerland}\\
\end{center}
\begin{abstract}
We explain the $\phi$ and $t$ dependences of
mesons with $J^{PC} = 0^{\pm +},1^{++},2^{\pm +}$ produced
in the central region
of $pp$ collisions.  For the $0^{++}$ and $2^{++}$ sector this reveals a
systematic behaviour in the data that appears to distinguish between
$q\bar{q}$ and non-$q\bar{q}$ or glueball candidates.
\end{abstract}
\newpage
\setcounter{footnote}{0}

The idea that glueball production might be favoured in the central region
of $pp \rightarrow pMp$ by the fusion of two
Pomerons (\pom ) is over twenty years old~\cite{robson,fcrpp}. The
fact that known $q\bar{q}$ states also are seen in this process frustrated
initial hopes that such experiments would prove to be a clean glueball
source. However, in \cite{ck97} we noted a kinematic effect whereby
known $q\bar{q}$ states could be suppressed leaving potential glueball
candidates more prominent. There has been an intensive experimental
programme in the last two years by the WA102 collaboration at CERN, which
has produced a large and detailed set of data on both the
$dP_T$
{}~\cite{ck97} and the azimuthal angle,
$\phi$,
dependence of meson
production\footnote{
$dP_T$ is the difference
in the transverse momentum vectors of the two exchange
Pomerons and
$\phi$ is the angle between the transverse momentum
vectors, $p_T$, of the two outgoing protons.
}.
\par
The azimuthal dependences
as a function of
$J^{PC}$ and the momentum transferred at the proton vertices , $t$,
are very striking.
As seen in refs.~\cite{WAphi,WA2-+}, and
later in this paper, the $\phi$ distributions for mesons with
$J^{PC}$~=~$0^{-+}$ maximise around $90^{o}$, $1^{++}$ at $180^o$ and $2^{-+}$
at $0^o$.
Recently, the WA102 collaboration has confirmed that this is not
simply a J-dependent effect~\cite{pi4papr} since
$0^{++}$ production peaks at $0^o$ for some states whereas others are more
evenly spread~\cite{WA0++}; $2^{++}$ established $q\bar{q}$ states peak at
$180^o$ whereas the $f_2(1950)$, whose mass may be consistent with the tensor
glueball predicted in lattice QCD, peaks at $0^o$~\cite{pi4papr}.

In this paper we show how these
phenomena arise and in turn expose the extent to which they could be
driven, at least in part, by the internal structure of the meson in
question and thereby be exploited as a glueball/$q\bar{q}$
filter~\cite{ck97}.
We find that the $\phi$ dependences of $0^{-+}$
and
$1^{++}$ follow on rather general grounds if a single trajectory
dominates the production mechanism. Having thus established the
ability to describe the phenomena quantitatively in these cases, we
predict the behaviour for
$2^{-+}$ production and then confront the $0^{++}$ and  $2^{++}$
$glueball/q\bar{q}$
sector.

\noindent $J^{PC}=0^{-+}$

Parity forbids the production of $0^{-+}$ by the fusion of two scalars
and also by the longitudinal ($``L"$) components of two vectors.
Transverse ($``T"$)  components are allowed and so we focus on the TT
component of \pom-\pom\thinspace  fusion in the production of the
$0^{-+}$ states.
$\rho \rho$ fusion is also possible, however, in this paper
we will concentrate on the $\eta^\prime$ meson
whose production has been found to be consistent with double
pomeron exchange~\cite{WAphi}.

The calculations have been described in~\cite{cs1,cs2} and the resulting
behaviour of the cross section may be summarised as follows:

\[
\frac{d\sigma}{dt_1 dt_2 d \phi^\prime}
\sim t_1 t_2 {G^p_E}^{2} (t_1)
{G^{p}_{E}}^2 (t_2) \sin^2(\phi^\prime) F^2(t_1, t_2, M^2)
\]
\noindent where $\phi^\prime$ is the angle between the two $pp$ scattering
planes in the \pom-\pom\thinspace centre of mass
and $F(t_1, t_2, M^2)$ is the \pom-\pom-$\eta^\prime$
form factor. We
temporarily set this equal to unity; $pp$ elastic scattering data
and/or a Donnachie Landshoff type form factor~\cite{dl} can be used as
model of the proton-\pom\thinspace form factor ($G_E^p(t)$).
This $\phi^\prime$ distribution is
shown in fig.~(1a) and applies in the meson rest frame (current-current
c.m.) in the ``symmetric" configuration, $t_1 = t_2; \vec{p}_{T1} = -
\vec{p}_{T2}; x_F = 0$. To generalise to real kinematics, we use
a Monte Carlo simulation based on Galuga
\cite{galuga} modified for $pp$ interactions and incorporating
the \pom-proton form factor from ref.~\cite{dl}. This
has the effect of distorting fig.~(1a) to fig.~(1b).


The WA102 collaboration measures the azimuthal angle ($\phi$)
in the $pp$ c.m. frame and so we
transform the $\phi^\prime$ from the current c.m. frame to $\phi$ for the
$pp$ c.m. frame. For the $0^{-+}$ case it happens that the above two
steps (fig.~(1a) to fig.~(1b) and this) tend to counterbalance.
Using the modified form of the Galuga Monte Carlo, discussed above,
we can now compare these predictions
with the experimental data, taking into account the experimental cuts
and the geometrical acceptance corrections of the WA102 experiment.
Any differences
between the output of the Monte Carlo model predictions
and the data are then due to intrinsic physics and not to
experimental acceptance effects.

In order to fit the data we found
that the \pom-\pom-meson form factor $ F(t_1, t_2, M^2) $
has to differ from unity.
If we
parametrise $ F^2(t_1, t_2, M^2) $
as $exp^{- b_T(t_1+t_2)}$ then we need $b_T$~=~0.5~$GeV^{-2}$
in order to describe the $t$
dependence.
Fig. (1c and 1d) compare the final theoretical form for the $\phi$
distribution and the $t$ dependence with the data for
the $\eta^\prime$.
The distributions are well described also for
the $\eta$ but it
has not yet been established that \pom-\pom \thinspace alone dominates the
production of this meson.

\noindent $J^{PC} = 1^{++}$

In refs.~\cite{cs1,cs2,fc1+} Close and Schuler have predicted that axial mesons
are produced polarised, dominantly in helicity one; this is verified by
data
\cite{WA1+pol}.  The cross section is predicted to have the form

\[
\frac{d\sigma}{dt_1dt_2d\phi^\prime} \sim t_1t_2 [ \{A(t_1^T,t_2^L) -
A(t_2^T,t_1^L)\}^2 + 4 A(t_1^T,t_2^L)A(t_1^L,t_2^T)\sin^2(\phi^\prime/2) ]
\]

\noindent where $A(t_i,t_j)$ are the \pom-\pom-$f_1$ form factors.
In the models of refs.~\cite{cs2,adln} the longitudinal Pomeron
amplitudes carry a factor of $1/\sqrt{t}$ arising from the fact that, in
the absence of any current conservation for the Pomeron, a longitudinal
vector polarisation is not compensated. Thus we make this factor explicit
and write $A(t_i,t_j^L) = \frac{\mu}{\sqrt{t_j}} a(t_i,t_j)$. The
cross section is predicted to behave as

\[
\frac{d\sigma}{dt_1dt_2d\phi^\prime} \sim
[\{\sqrt{t_1} - \sqrt{t_2}\frac{a(t_1^T,t_2^L)}{a(t_1^L,t_2^T)}\}^2 + 4
\sqrt{t_1t_2} \frac{a(t_1^T,t_2^L)}{a(t_1^L,t_2^T)}
\sin^2(\phi^\prime/2)]a^2(t_1^L,t_2^T)
\]

In the particular case where the ratio of form factors is unity, this
recovers the form used in ref.~\cite{cs2}

\[
\frac{d\sigma}{dt_1dt_2d\phi^\prime} \sim
[(\sqrt{t_1} - \sqrt{t_2})^2 + 4 \sqrt{t_1t_2}
\sin^2(\phi^\prime/2)]a^2(t_1,t_2)
\]

\noindent which implies a dominant $\sin^2(\phi/2)$ behaviour that tends to
isotropy when suitable cuts on $t_i$ are made. This is qualitatively
realised (figs.~1e and f of ref.~\cite{WAphi}).

We have
parametrised
$a(t_i^T,t_j^L)$
as an exponential,
$exp^{-(b_Tt_i+b_Lt_j))}$ where $i,j=1,2$; $b_T$~=~0.5 $GeV^{-2}$ was
determined from the
$\eta^\prime$ data above; $b_L$ is
determined from
the overall $t$ dependence of the $1^{++}$ production and requires
$b_L$~=~3~$GeV^{-2}$.
Fig.~(2a and b) show the output of the model predictions from the
Galuga Monte Carlo superimposed
on the $\phi$ and $t$ distributions for the $f_1(1285)$ from
the WA102 experiment.

In addition we have a parameter free
prediction of
the variation of the $\phi$ distribution as a function of
$|t_1-t_2|$.
Fig.~(2c and d) show the output of the Galuga Monte Carlo superimposed
on the $\phi$ for the $f_1(1285)$ for $|t_1 - t_2|$~$\le$~0.2~$GeV^{-2}$
and $|t_1 - t_2|$~$\ge$~0.4~$GeV^{-2}$ respectively.
The agreement between the data and
our prediction is excellent.
Similar conclusions arise for the $f_1(1420)$.

\noindent $J^{PC}=2^{-+}$

The $J^{PC}$~=~$2^{-+}$ states, the $\eta_2(1645)$ and $\eta_2(1870)$,
are predicted to be produced
polarised. Helicity 2 is suppressed by Bose symmetry~\cite{cs1} and
has been found to be negligible experimentally~\cite{WA2-+}.
The structure of the cross section
is then predicted to be

\noindent (i) helicity zero:  as for the $0^{-+}$ case,
\[
\frac{d\sigma}{dt_1dt_2d\phi^\prime} \sim t_1t_2 \sin^2(\phi^\prime)
\]

\noindent (ii) helicity one:

\[
\frac{d\sigma}{dt_1dt_2d\phi^\prime} \sim
[\{\sqrt{t_1} - \sqrt{t_2}\frac{a(t_1^T,t_2^L)}{a(t_1^L,t_2^T)}\}^2 + 4
\sqrt{t_1t_2} \frac{a(t_1^T,t_2^L)}{a(t_1^L,t_2^T)}
\cos^2(\phi^\prime/2)]a^2(t_1^L,t_2^T)
\]
\noindent which is as the $1^{++}$ case except for the important and
significant change from $\sin^2(\phi^\prime/2)$ to $\cos^2(\phi^\prime/2)$.

The intrinsic relative strengths of the two helicity amplitudes will
depend upon the internal dynamics of the \pom-\pom-$\eta_2$ vertex
which are beyond the scope of the present paper. The
uncompensated factor of $t_1t_2$ in the helicity zero component will
tend to suppress this kinematically under the conditions of the WA102
experiment. Indeed, WA102 find that helicity one alone is able to
describe their data~\cite{WA2-+}; this is in interesting contrast to
$\gamma
\gamma
\to \eta_2(Q\bar{Q})$ in the non-relativistic quark model where the
helicity-one amplitude would be predicted to vanish~\cite{fczpli}. We shall
concentrate on this helicity-one amplitude henceforth.

The results of the WA102 collaboration for the $\eta_2(1645)$~\cite{WA2-+}
are shown
in fig.~(3a and b).
The distribution peaks as $\phi \to 0$, in
marked contrast to the suppression in the $1^{++}$ case (fig.~2a).

Integrating our formula over $\phi$, with the same approximations as
previously, implies

\[
\frac{d\sigma}{dt_1dt_2} \sim (t_1 + t_2) ( exp^{-(b(t_1+t_2)})
\]

\noindent and, in turn, that

\begin{equation}
\frac{d\sigma}{dt} \sim (1 + bt) ( exp^{-bt})
\label{eq:a}
\end{equation}

\noindent This simple form compares remarkably well with WA102 who fit to
$\alpha e^{-b_1t} + \beta t e^{-b_2t}$; our prediction (eq.~\ref{eq:a}) implies
that $b_1 \equiv b_2$ and that $\beta/\alpha \equiv b $ and WA102
find for the $\eta_2(1645)$)~\cite{WA2-+}
$b_1 = 6.4 \pm 2.0  ; b_2 = 7.3 \pm 1.3 $ and $\beta =2.6 \pm 0.9$,
$\alpha = 0.4 \pm 0.1$

Performing the
detailed comparison of model and data via Galuga, as in
the previous examples, leads to the results shown in fig.~(3a and b) for
the $\eta_2(1645)$; the $\eta_2(1870)$ results are qualitatively similar.
Bearing in mind that there are no free parameters, the
agreement is remarkable. Indeed, the successful
description of the $0^{-+}$, $1^{++}$ and $2^{-+}$ sectors, both
qualitatively and in detail, set the scene for our
analysis of the $0^{++}$ and $2^{++}$ sectors where glueballs are
predicted to be present together with established $q\bar{q}$ states.
Any differences between data and
this model may then be
a signal for hadron structure, and potentially a filter for glue degrees of
freedom.

\noindent $J^{PC}=0^{++}$ and $2^{++}$

In contrast to the $0^{-+}$ case, where parity forbade the LL amplitude,
in the $0^{++}$ case both $TT$ and $LL$ can occur. Hence there are two
independent form factors~\cite{cfl} $A_{TT}(t_1,t_2,M^2)$ and
$A_{LL}(t_1,t_2,M^2)$. For
$0^{++}$ and the helicity zero amplitude of $2^{++}$ (which
experimentally is found to dominate~\cite{WAhel0}) the angular dependence
of scalar meson production will be~\cite{cs2}

\begin{equation}
\frac{d\sigma}{dt_1 dt_2 d \phi^\prime}
\sim {G^p_E}^{2} (t_1)
{G^{p}_{E}}^2 (t_2)[1 +
\frac{\sqrt{t_1t_2}}{\mu^2}\frac{a_T}{a_L}e^{(b_L-b_T)(t_1+t_2)/2 }
\cos(\phi^\prime)]^2 e^{-b_L(t_1+t_2) }
\label{eq:b}
\end{equation}

\noindent where we have written $a_L(t) = a_Le^{-(b_Lt/2)}$ and
$a_T(t) = a_Te^{-(b_Tt/2)}$ with $b_{L,T}$ fixed
to the values found earlier. The ratio
$a_T/a_L$ can be positive or negative, or in general even complex.

  Eq.(\ref{eq:b}) predicts that there should be significant changes
in the $\phi$ distributions as $t$ varies.
When
$\frac{\sqrt{t_1t_2}}{\mu^2}a_T/a_L \sim \pm1$, the $\phi$ distribution will be
$\sim \cos^4(\frac{\phi}{2})$ or $\sin^4(\frac{\phi}{2})$
depending on the sign.
Indeed data on the
enigmatic scalars $f_0(980)$ and $f_0(1500)$ show a
$\cos^4(\frac{\phi}{2})$ behaviour when
$\sqrt{t_1t_2} \leq 0.1$ GeV$^2$,
changing to $\sim \cos^2(\phi)$ when $\sqrt{t_1t_2} \geq
0.3$ GeV$^2$~\cite{WAphi}.

In this paper we show how the overall $\phi$
dependences for the $f_0(1370)$, $f_0(1500)$, $f_2(1270)$ and
$f_2(1950)$ can be described by varying the quantity $\mu^2a_L/a_T$.
Results are shown in fig.~4.
It is clear that these $\phi$
dependences discriminate two classes of meson in the $0^{++}$ sector and
also in the $2^{++}$.
The $f_0(1370)$ can be described using $\mu^2a_L/a_T$~=~-0.5~$GeV^2$,
for the $f_0(1500)$ it is +0.7~$GeV^2$,
for the $f_2(1270)$ it is -0.4~$GeV^2$ and
for the $f_2(1950)$ it is +0.7~$GeV^2$.

It is interesting to note that we can fit these $\phi$ distributions
with one parameter and it is primarily the sign of this quantity that
drives the $\phi$ dependences.
Understanding the dynamical origin of this sign is now a central issue
in the quest to
distinguish $q \overline q$ states from
glueball or other exotic states.

In summary, for the production of $J^{PC}$~=~$0^{-+}$ mesons
we can predict the $\phi$ dependence and
the vanishing cross section as $t \rightarrow 0$ absolutely and fit the
$t$ slope in terms of one parameter,
$b_T$.
For the $J^{PC}$~=~$1^{++}$ mesons
we predict the general form for the $\phi$
distribution. By fitting the $t$ slope we obtain the parameter
$b_L$; this then gives a parameter free prediction for
the variation of the $\phi$ distribution as a function of $t$ which
agrees with the data.
In addition, these give
absolute predictions for the $t$ and $\phi$ dependences of the
$J^{PC}$~=~$2^{-+}$ mesons which are again in accord with the data
when helicity 1 dominance is imposed.
For the $0^{++}$ and $2^{++}$ sector we extract a systematic
behaviour from the data that requires a dynamical interpretation. Whether
this is the long sought discriminator between $q\bar{q}$ and
non-$q\bar{q}$ states is for the future.

\begin{center}
{\bf Acknowledgements}
\end{center}
\par
This work is supported, in part, by grants from
the British Particle Physics and Astronomy Research Council,
the British Royal Society,
the European Community Human Mobility Program Eurodafne,
contract NCT98-0169 and
the EU Fourth Framework Programme contract FMRX-CT98-0194.

\clearpage
{ \large \bf Figures \rm}
\begin{figure}[h]
\caption{
The predicted $\phi^\prime$ distributions for $J^{PC}$~=~$0^{-+}$ mesons
a) naive distribution and b) taking into account the experimental kinematics.
c) The $\phi$
and d) the $|t|$ distributions for the $\eta^\prime$
for the data (dots) and the model predictions from the Monte Carlo (histogram).
}
\label{fi:1}
\end{figure}
\begin{figure}[h]
\caption{
a) The $\phi$
and b) the $|t|$
distributions for the $f_1(1285)$
for the data (dots) and the Monte Carlo (histogram).
c) and d) the $\phi$ distributions for $|t_1 - t_2|$~$\le$~0.2 and
$|t_1 - t_2|$~$\ge$~0.4~$GeV^{2}$ respectively.
}
\label{fi:2}
\end{figure}
\begin{figure}[h]
\caption{
a) The $\phi$
and b) the $|t|$
distributions for the $\eta_2(1645)$
for the data (dots) and the Monte Carlo (histogram).
}
\label{fi:3}
\end{figure}
\begin{figure}[h]
\caption{
The $\phi$
distributions for the a) $f_0(1370)$,
b) $f_0(1500)$, c) $f_2(1270)$ and d) $f_2(1950)$
for the data (dots) and the Monte Carlo (histogram).
}
\label{fi:4}
\end{figure}
\begin{center}
\epsfig{figure=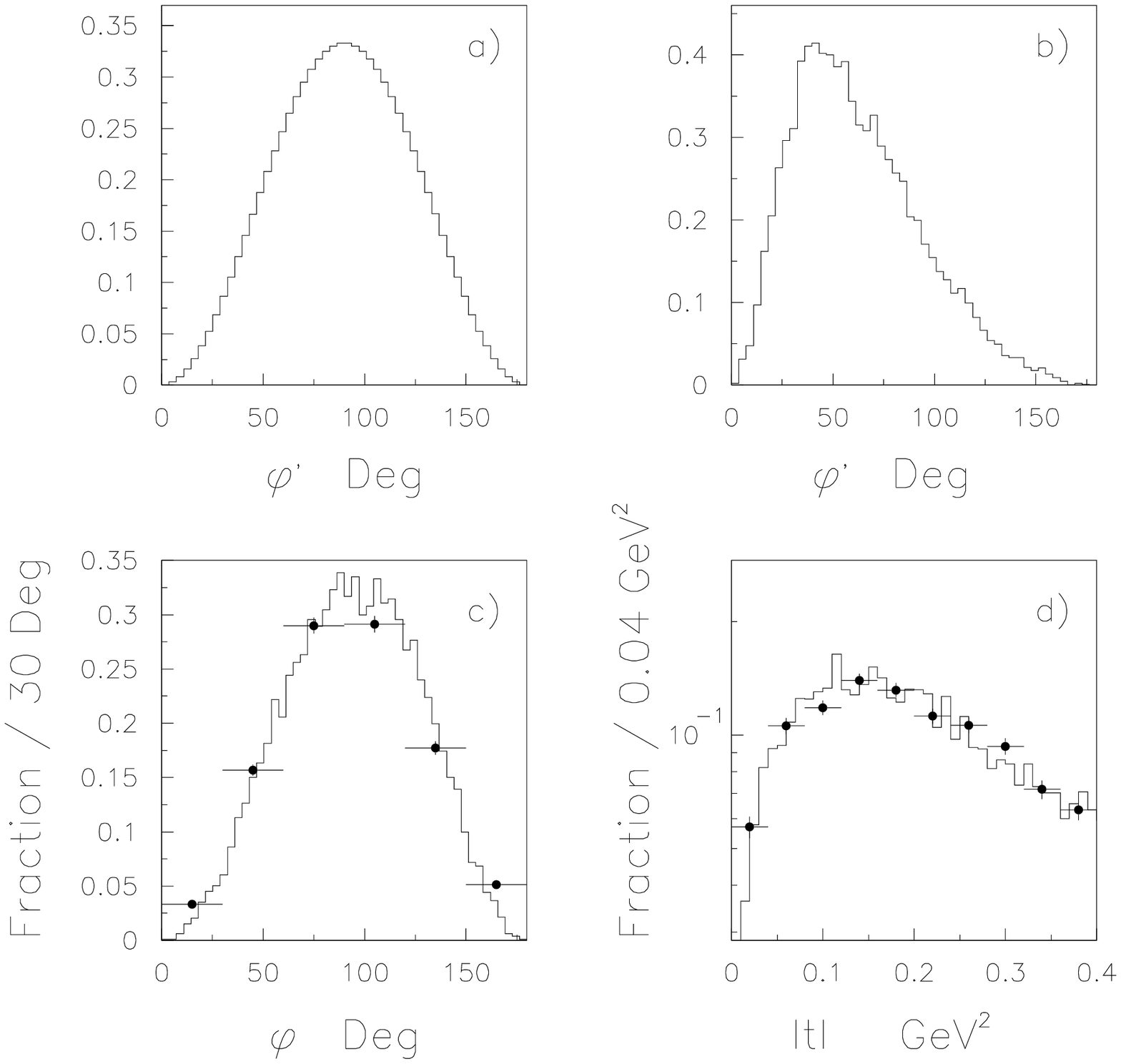,height=22cm,width=17cm}
\end{center}
\begin{center} {Figure 1} \end{center}
\newpage
\begin{center}
\epsfig{figure=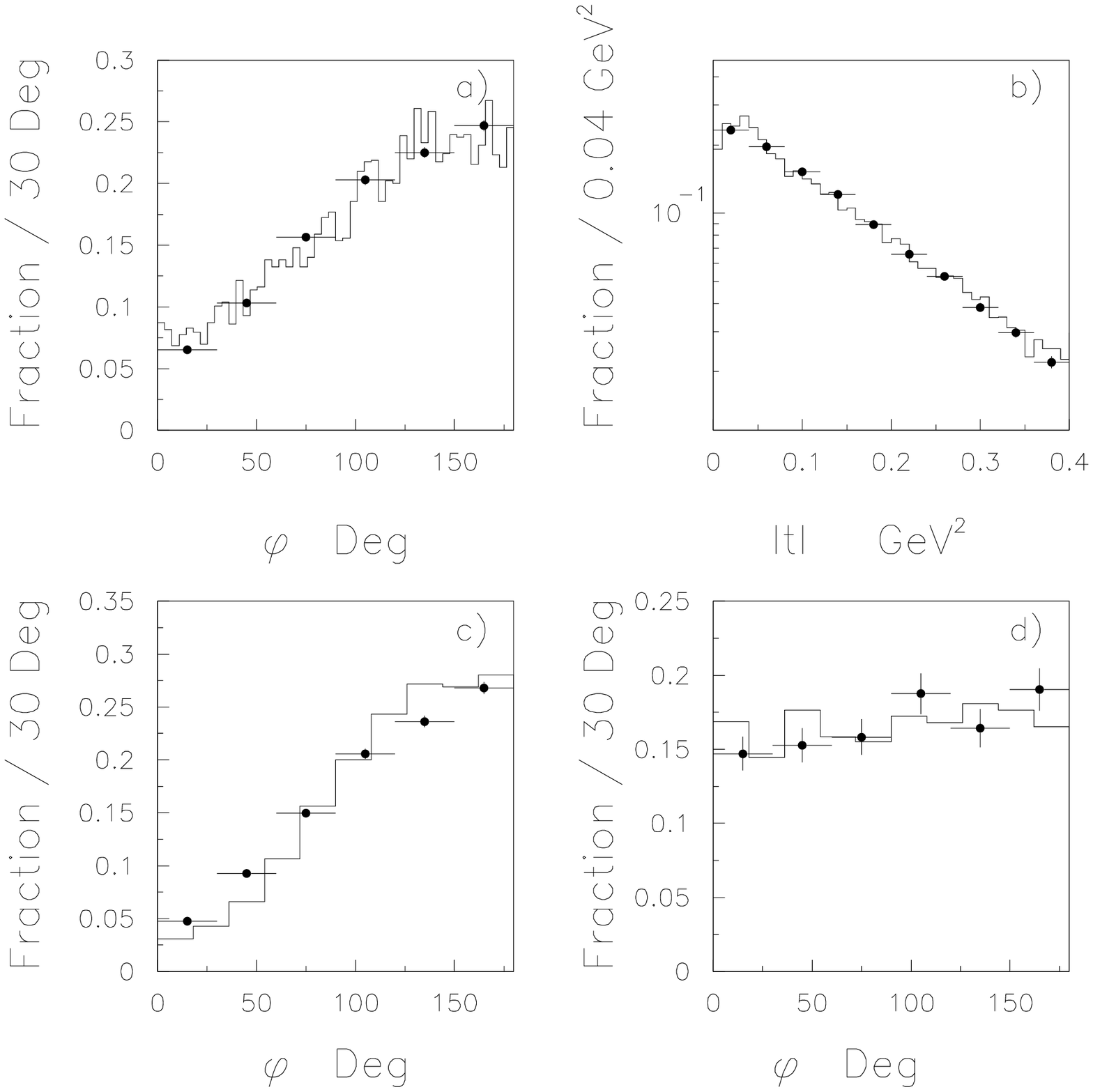,height=22cm,width=17cm}
\end{center}
\begin{center} {Figure 2} \end{center}
\newpage
\begin{center}
\epsfig{figure=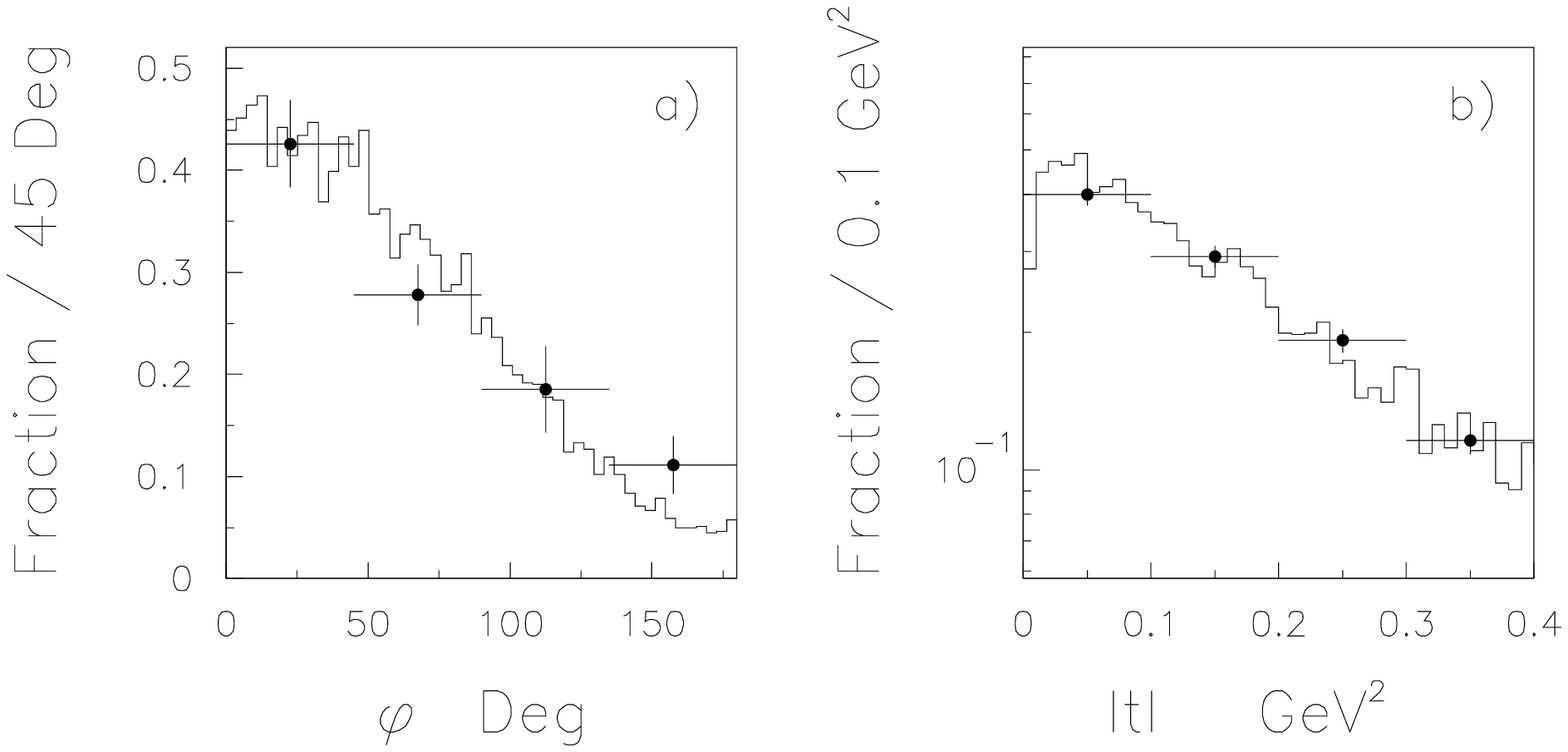,height=22cm,width=17cm}
\end{center}
\begin{center} {Figure 3} \end{center}
\newpage
\begin{center}
\epsfig{figure=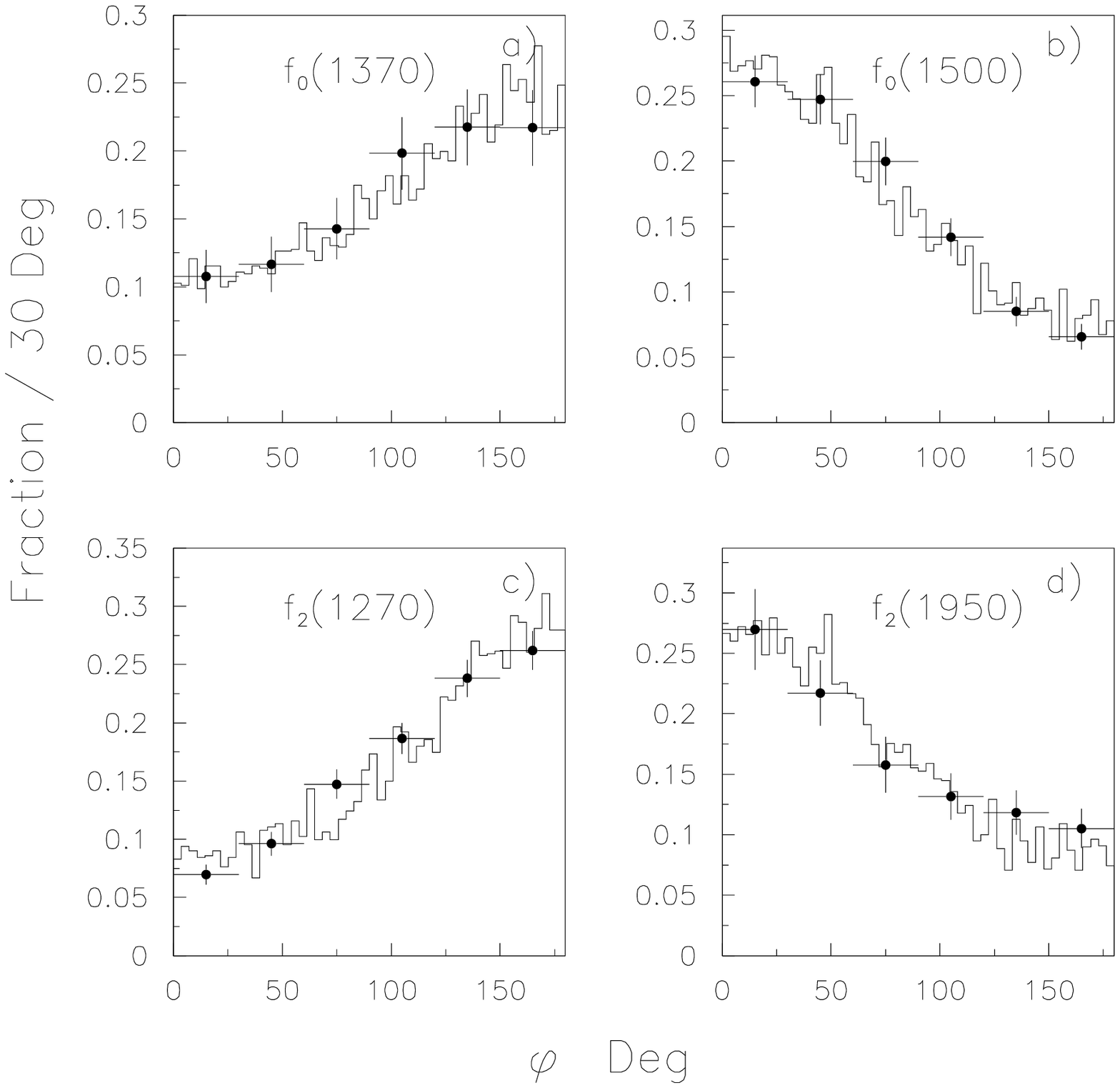,height=22cm,width=17cm}
\end{center}
\begin{center} {Figure 4} \end{center}
\end{document}